\documentclass[prl,showpacs,twocolumn]{revtex4} 
\usepackage{hyperref}
\usepackage[dvips]{graphicx,color}
\begin{document}
\preprint{APS/123-QED}
\title{Regular and irregular 
modulation of frequencies
in limit cycle oscillator networks}
\author{Masashi Tachikawa}
\email{mtach@complex.c.u-tokyo.ac.jp}
\affiliation{
ERATO Complex Systems Biology Project, 
           JST, 3-8-1, Komaba, Meguro-ku, Tokyo 153-8902, Japan.}
\date{\today}

\begin{abstract} 
Frequency modulation by perturbation is the essential trait
that differentiates 
limit cycle oscillators from phase oscillators.
We studied networks of identical limit cycle oscillators whose
 frequencies are modulated sensitively by the change of their
 amplitudes, and demonstrated that the frequencies sustainably take  
distributed values.
We observed two complex phenomena in the networks: 
stationarily 
distributed frequencies at a regular interval, 
and continuous irregular modulation of frequencies. 
In the analysis we reveal the mechanisms 
by which 
the frequencies are distributed, and
show how the sensitive modulation of frequencies 
produces 
these complex phenomena. 
We also illustrate the topology of networks regulating 
the behaviors of the systems.
\end{abstract} 


\maketitle


Oscillator networks 
are widely accepted as a standard model 
for understanding 
dynamical systems with
many degree of freedom, 
because of the accumulation of 
knowledge about them 
\cite{Win67,Kura84,Piko03,KT01} and 
the large number  
of their applications \cite{Wie96}.  
The statistical synchrony for simple phase oscillator 
networks has been well investigated 
since the celebrated 
works by Winflee \cite{Win67} and  
Kuramoto \cite{Kura84}.
Recently this field has been developed by incorporating 
statistical 
knowledge about complex networks \cite{collect}. 
At the same time, 
detailed dynamic behaviors are also studied 
for oscillators showing high dimensional properties.
Various types of coherent motions 
are classified, and concepts 
such as  
phase synchronization \cite{Piko03},  
clustering, and 
chaotic itinerancy \cite{KT01} 
have been 
proposed. 
However, a large number 
of complex phenomena 
still remain unrevealed, and the integrated understanding 
about them is still lacking 
compared with 
that about 
simple phase oscillator systems.
Which property of a high-dimensional 
oscillator 
is essential for generating 
complex phenomena, and how does that differ from the functioning of a
phase oscillator? 
Among others, the change of 
{\it frequency} (phase evolution speed) should be emphasized.
While any perturbation on a phase oscillator 
produces only a phase shift, 
the frequency 
in a high-dimensional oscillator can be modulated 
\footnote{
In other words, if effects of perturbations on a limit cycle oscillator
are described by phase shifts, the system (oscillator + perturbations)
is described by a phase oscillator system \cite{Kura84}.}.
When such oscillators are assembled to form a network, 
the resulting system  
can generate distributed frequencies states 
and lead to the temporal formation and 
breakdown of synchrony of the system. 
These behaviors are observed in many complex systems
\cite{KT01,BD04,Kay05,Shi95,Tak97}, and in fact 
variable frequency effects have been noted 
in some them.
Several theoretical studies have pointed to 
the variable frequency effect on complex systems 
\cite{Tak97,Dai06,RN06,TF07}.

In the present paper, 
we introduce networks of interacting limit cycle oscillators, 
whose frequencies are modulated by changes in their amplitudes, 
and demonstrate that amplitudes of oscillators spontaneously 
take distributed values; i.e. the frequencies are distributed.
Additionally, we observe that there are two types of behaviors  
 depending on network topologies: 
arrangement of frequencies at a regular interval 
and continuous irregular modulation of frequencies.
The analysis reveals the mechanism that 
generates distributed frequencies states. We also show 
how the topology of the networks regulates the behaviors of the system.


Let us consider a limit cycle oscillator that is weakly stable in one 
direction transverse to the rotation direction. 
If one produces perturbations of intensity 
comparable to the attraction of the limit cycle,  
the orbit widely deviates from the limit cycle, 
introducing considerable modulation into the frequency.
In order to represent this effect in a simple model, 
we introduce a two-variable oscillator by assuming that 
the deviations in transverse directions are 
projected in a one-dimensional variable $r$ called {\it amplitude}. 
Using $\theta$ as the phase variable defined along with the 
limit cycle, and taking the first order of $r$ into account, 
we get the model equations, 
\begin{eqnarray} 
\begin{array}{l} 
\displaystyle \dot{\theta} = \omega_0 + \Omega(\theta) r    \\
\displaystyle \dot{r} = - \Gamma(\theta)  r 
\end{array} 
+~{\rm (perturbation)}.
\end{eqnarray}
$\Omega(\theta)$ represents the modulation of the frequency  
by the amplitude change, and $\Gamma(\theta)$ indicates the 
decay rate of $r$ which is assumed to be small. 
If the phase evolution is sufficiently faster than the change in the 
amplitude (weak stability assumption), 
the phase dependence of $\Gamma(\theta)$ 
is normalized and replaced with 
the parameter $\gamma$. 
We adopt $\omega_0=1$, $\Omega(\theta)=
\omega\cdot(1+\varepsilon\sin(2\pi\theta))$
in this study.

\begin{figure*}
\includegraphics[width=8.5cm]{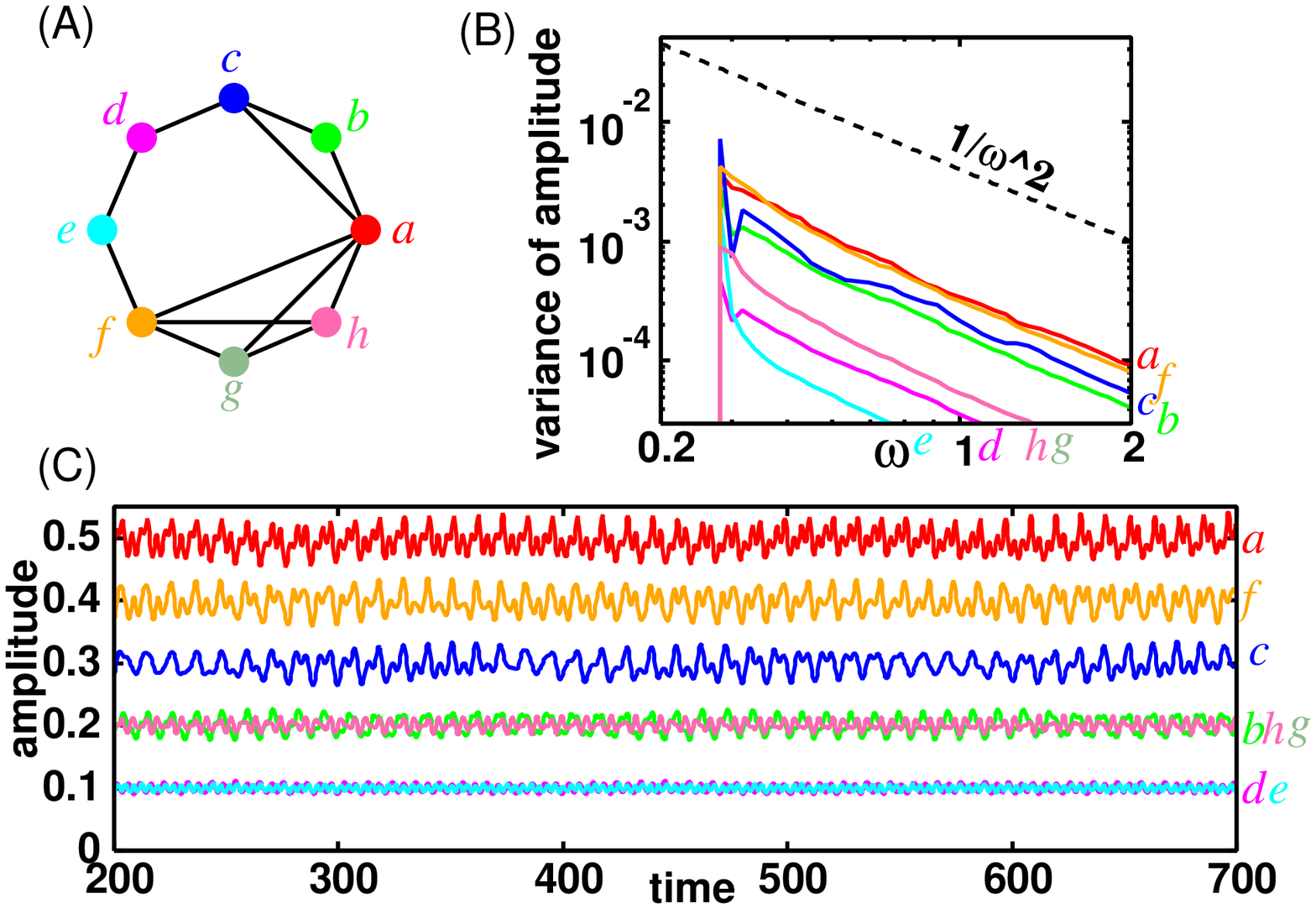}
\includegraphics[width=8.5cm]{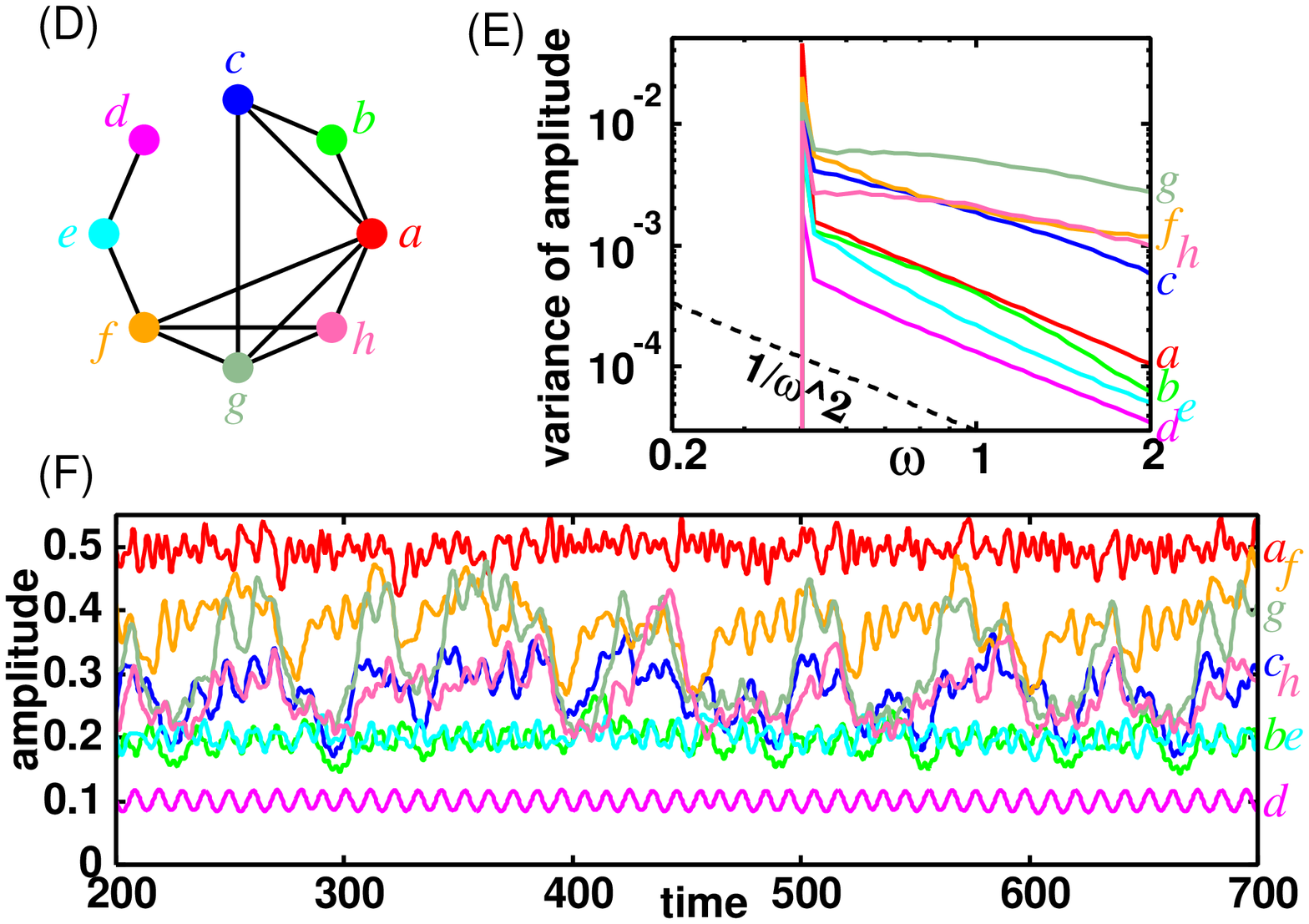}
\caption{ 
(color online) Behaviors of coupled oscillator networks indicated by 
Fig. A  and D are shown at Fig. B, C and E, F, respectively.
$\varepsilon=0.5$, 
in B, C, E and F, 
and $\omega=1$  in C and F. 
B and E show the variances of amplitudes of oscillators in network A
and D as functions of $\omega$. 
Dashed lines indicate the slope of $1/\omega^2$. 
C and F show the time series of amplitudes of oscillators 
in network A and D. 
}
\label{fig:1}
\end{figure*}

Next, we introduce the oscillator networks. 
Oscillators are supposed to interact 
in a bidirectional way 
with an interaction matrix $k_{ij}=k_{ji}\in\{0,1 \}$. 
Note that the phase differences between interacting oscillators 
are not always small, and large differences in phases 
produce considerable perturbations in the amplitudes. 
We select the interaction function to agree with the diffusive 
coupling at the leading order, which satisfies 
$\partial \dot{\theta}/\partial\Delta\theta|_{\Delta r =0}  \neq0$, 
$\partial \dot{r}/\partial\Delta\theta|_{\Delta r =0}  =0$, and 
$\partial^2 \dot{r}/\partial\Delta\theta^2|_{\Delta r =0}  \neq0$, 
where $\Delta \theta$ and $\Delta r$ are phase differences 
and amplitude differences between the oscillators. 
Thus, the models of the oscillator networks 
are given by 
\begin{eqnarray} 
\displaystyle \dot{\theta}_{\alpha} &=&  1+
\omega\left(1+\varepsilon\sin(2\pi\theta_{\alpha})\right) r_{\alpha}
\nonumber  \\
\displaystyle 
&&~~~~~~~~~~~~+  
D_{\theta} \sum_{\beta} k_{\alpha\beta}
\sin\left( 2\pi(\theta_{\beta}-\theta_{\alpha}) \right),
\label{eq:model_1}   \\
\displaystyle \dot{r}_{\alpha} &=& 
- \gamma r_{\alpha} 
+ D_{r} \sum_{\beta} k_{\alpha\beta}
\left\{1-\cos\left( 2\pi(\theta_{\beta}-\theta_{\alpha}) 
\right)\right\}, \label{eq:model_2}  
\end{eqnarray}  
where $\alpha$, $\beta$ are oscillator indexes.
In this study, 
we set 
$D_{\theta}=D_{r}=0.01, \gamma=0.1$, $\varepsilon=0, 0.5$
and  changed the control parameter $\omega=0.2\sim 2$,
which indicates the magnitude of the modulation of the frequency 
by the amplitude change. 
All simulations run from uniformly distributed 
random initial conditions in $r_{\alpha}\in (0,1],~
\theta_{\alpha}\in(0,1]$.


Now we take up two examples of networks 
indicated by Fig. \ref{fig:1}-A and Fig. \ref{fig:1}-D 
and demonstrate two distinct behaviors of 
the oscillator-networks.

Figure \ref{fig:1}-C shows the time evolution of amplitudes 
of oscillators in network A (Fig. \ref{fig:1}-A) 
after they reach the stationary states.
The case with $\omega=1$, $\varepsilon=0.5$ is shown. 
In the figure, 
amplitudes are distributed and remain around the discrete stationary points 
which are arranged at regular intervals of $0.1$. 
The fluctuations of amplitudes around the stationary points 
are small and seem to be periodic or quasi-periodic.
In other words, we clearly observe the layers of amplitudes. 
The variances of the amplitudes are plotted 
against $\omega$ in Fig. \ref{fig:1}-B. 
With $\omega \le 0.38$,  
all oscillators  achieve complete synchronization 
($r_{\alpha}=0,\theta_{\alpha}=\theta_{\beta}$  for all
$\alpha,\beta$), 
and  no fluctuation appears \footnote{
In this model, 
the complete synchronization state is always stable.
However, with large $\omega$ values, 
this state is not reached from most initial conditions.
}. 
With $\omega > 0.38$, amplitudes are distributed and 
 the averaged values of amplitudes are independent 
from $\omega$ (data not shown).
In the region, 
the variances of amplitudes decay with  $1/\omega^2$.

On the other hand, network D (Fig. \ref{fig:1}-D) 
exhibits 
a different behavior as shown in Fig. \ref{fig:1}-E and -F.
At the steady state, 
the layers of amplitudes are partially broken and 
the amplitudes of oscillators $c$, $f$, $g$  and $h$ show large fluctuations.
The widths of their fluctuations are greater than the range between 
layers, 
and the fluctuations decay slowly with the increase of $\omega$.
Other oscillators ($a$, $b$, $d$ and $e$) still stay around the 
stationary points.
We emphasize that network D is  constructed by changing only one 
interaction from network A: $c$-$d$ to $c$-$g$.
Thus, behaviors are quite sensitive to 
the topology of networks.

\begin{figure}
\includegraphics[width=6.5cm]{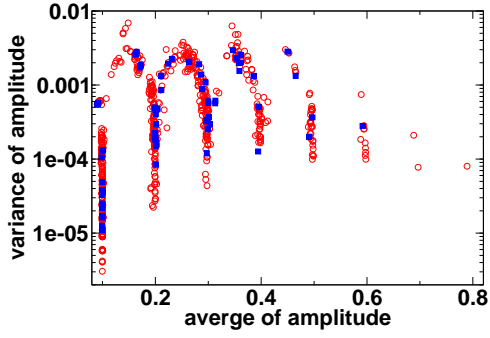}
\caption{
(color online)
Scatter plot distributions of 
 averages and variances of amplitudes of all oscillators 
in two large networks with 100 and 1000 oscillators.
The networks are generated randomly with connection probabilities 
 $2/$(number of oscillators).
$\varepsilon=0.5$, 
 $\omega=1$.
Data of the 1000-oscillator network are plotted with red open circles, 
and data of the 100-oscillator network are plotted with blue filled 
squares.
}
\label{fig:3}
\end{figure}

Summarizing the results, 
the amplitudes of oscillators in networks are widely distributed  
when the frequency modulation effect ($\omega$) is large. 
We also found two distinct behaviors in amplitudes
depending on the topologies of 
networks:
(i) amplitudes of all oscillators in network form 
layers, and 
(ii) amplitudes of a part of oscillators show 
 large fluctuations across the layers.
Type (i) behavior 
is less frequently observed in larger networks.
When $1000$ networks were generated with $10$ and $50$ oscillators,
respectively, $468$ networks with $10$ oscillators showed type (i)
behavior and only $25$ networks with $50$ oscillators showed
type (i) behavior\footnote{
$1000$ networks for each size 
were generated randomly with connection probabilities
$2/$(number of oscillators).
$\varepsilon=0.5$, 
and $\omega=1$ 
were used to test the behaviors.
}.
Meanwhile, both widely fluctuated and less fluctuated oscillators 
always coexist in large networks. 
Figure  \ref{fig:3} shows scatter plot distributions of 
the averages of amplitudes and the variances of amplitudes 
for two random networks which have 100 and 1000 oscillators.
The variances are widely distributed.
For less variant oscillators, the averaged amplitudes show 
the order in regular intervals as seen in Fig. \ref{fig:1}-C. 

In the following, we illustrate the mechanisms of 
generating both behaviors. 
In the analysis $\varepsilon=0$ is used for simplicity.

We begin with investigating the mechanism of  
 forming layers. 
In order for the amplitudes to take stationary values, 
interaction terms should be approximately constant.
There are two possible mechanisms; 
one is phase locking and the other is averaging.
If the phase difference between two interacting oscillators takes a constant
value (a phase-locked state), the interaction term also becomes constant.
However, it is unlikely that the phase differences between all interacting
oscillators take constant values except in complete synchronization,
 where all oscillators assume 
an identical state 
($r_{\alpha}=0,\theta_{\alpha}=\theta_{\beta}$  for all
$\alpha,\beta$). 
On the other hand, 
the averaging assumption for an interaction holds 
when the two oscillators have considerably different frequencies so that 
the time evolution of amplitudes can not follow the alternation
of the interaction term. 
Since the difference in amplitudes produces 
a difference in frequencies,  
the averaging assumption is expected to hold.

Assume that the speeds of the phase differences 
between all coupled oscillators are sufficiently fast that 
interaction functions in eq. (\ref{eq:model_2})
are approximately 
replaced with the averaged values 
($\cos\left( 2\pi(\theta_{\beta}-\theta_{\alpha}) 
\right)\to 0$). 
Then, eq. (\ref{eq:model_2}) is solved and 
the approximate amplitude takes 
$N_{\alpha}D_{r}/\gamma$,
where $N_{\alpha}$ is the number of interactions for oscillator $\alpha$.
Note that the approximate amplitudes  are proportional to 
$N_{\alpha}$, and hence they take discrete values.
With the parameters used in Fig. \ref{fig:1}-C, 
$D_{r}/\gamma=0.1$ which agrees with the observed interval in the figure.

Here we examine under which conditions 
the averaging assumption holds.
The following equation 
expresses the phase difference between two interacting oscillators 
in which amplitudes are replaced with approximate values 
\footnote{
Here, the amplitudes are replaced with approximate values 
and the interaction in phases is taken into account.
This is because the amplitudes are supposed to be {\it slow} variables.
}:
\begin{eqnarray}
\dot{\overline{\Delta\theta}}_{\alpha \beta}= 
\left(N_{\alpha}-N_{\beta} \right)\frac{D_{r}\omega}{\gamma}
 -  D_{\theta}\sin\left(2\pi\overline{\Delta\theta}_{\alpha \beta} \right). 
\label{eq:pdif}
\end{eqnarray}
We also assume that the interaction effects from other 
oscillators are replaced by averaged values.
If $|N_{\alpha}-N_{\beta}|>0$ and $D_{r}\omega/(\gamma D_{\theta}) \gg 1$ 
hold, the interaction has little effect and  
a uniform oscillation with frequency 
$F_{\alpha\beta}=D_{r}(N_{\alpha}-N_{\beta})\omega/\gamma$
is an approximate solution of eq. (\ref{eq:pdif}). 
Putting the solution
 in eq (\ref{eq:model_2}), we get the equation for 
the amplitude 
\begin{eqnarray}
\dot{\overline{r}}_{\alpha} =  -\gamma \overline{r}_{\alpha}
 + D_{r}(N_{\alpha}-1) 
+D_{r}\left\{1-\cos\left(2\pi \overline{\Delta\theta}_{\alpha\beta} 
\right)\right\}, 
\label{eq:rd_amp} 
\end{eqnarray} 
where the effects from other interactions are again replaced 
by averaged values. 
The solution is 
\begin{eqnarray} 
\overline{r}_{\alpha}=   \frac{D_{r} N_{\alpha}}{\gamma} + 
\frac{D_{r}}{ \sqrt{(2\pi F_{\alpha\beta})^2+\gamma^2}}\sin( 
2\pi F_{\alpha\beta}t +\theta_0),  \label{eq:solv}
\end{eqnarray}
where $\theta_0$ is the initial phase.
The second term denotes that the variance of the  amplitude
decreases along with the increase of $\omega/\gamma$.
This is also consistent with the decrease of 
fluctuation in Fig. \ref{fig:1}-B, since 
$1/\sqrt{\rm variance} \propto F_{\alpha\beta} \propto \omega$.  
Thus, the averaging assumption is fulfilled.

When $N_{\alpha} = N_{\beta}$ holds, 
the averaging assumption does not hold.
The synchronous state ($\overline{\Delta\theta}_{\alpha\beta}=0$)
 becomes the attractor for eq. (\ref{eq:pdif}).
Thus, we get 
$\overline{r}_{\alpha}=\overline{r}_{\beta} = (N_{\alpha}-1)D_{r}/\gamma$
from eq. (\ref{eq:rd_amp}).
Although this means that the assumed relation for amplitudes 
($\overline{r}_{\alpha}=\overline{r}_{\beta}=N_{\alpha}D_{r}/\gamma$) 
in eq. (\ref{eq:pdif}) is violated, 
both amplitudes change simultaneously and 
the relation $r_{\alpha}=r_{\beta}$ still holds.
Hence, the synchrony of oscillators $\alpha$ and $\beta$  
is sustained as a special phase-locked case.

To sum them up, two interacting oscillators with different
$N_{\alpha}$ have sufficient difference in 
frequency to be averaged, 
and the interaction is approximately replaced by the averaged value.
On the other hand, two oscillators with the same $N_{\alpha}$
have the same frequency on average, and they synchronize with 
each other, 
an interaction that produces little effect on the amplitudes. 
Let us apply the analysis to network A
and check the relation between the topology of the network and 
the observed layers in Fig. \ref{fig:1}-C.
In network A, $d$-$e$ and $g$-$h$ pairs have the same numbers of 
interactions
($N_d=N_e=2$, $N_g=N_h=3$); thus, they are supposed 
to be synchronizing pairs.
Therefore, the averaged amplitudes of oscillators are given by 
\begin{eqnarray}
\overline{r}_{\alpha}=\frac{N'_{\alpha}D_{r}}{\gamma},
~~~~ \alpha=a,\cdots,h
\label{eq:approx_r} 
\end{eqnarray}
where $N'_{\alpha}$ denotes the number of effective interactions 
after eliminating the interaction of 
synchronizing pairs.
They show good agreement 
with the order of amplitudes  
in regular intervals as in Fig. \ref{fig:1}-C. 
Besides, the phase difference between oscillators $d$ and $e$ 
is always less than $0.013$  
and oscillators $g$ and $h$ have exactly the same phase 
in the simulations.

Now we investigate the continuous irregular motion  
of amplitudes observed in the network D.
In short, it is caused by the indeterminacy of a consistent set of 
synchronizing pairs.
This leads to continuous changes of the synchronizing pairs,
and irregular motion arises.
Here we illustrate these processes in detail using network D 
as an example.

Following the above rules, 
oscillators $f$ and $g$ in network D 
 synchronize ($N_f=N_g=4$), and 
their amplitudes converge  to $3D_{r}/\gamma$.
Additionally, oscillator $c$ and $h$,which interact with $g$, 
have the same frequency, since $N_c=N_h=N_g-1=3$.
Thus, $g$ and either of two oscillators start to synchronize, 
which brings about another change in amplitudes of the synchronizing pair 
($c$-$g$ or $g$-$h$). 
Their amplitudes converge to $2D_{r}/\gamma$. 
However, this introduces a  
difference in amplitude between oscillator
$f$ and $g$, i.e. a difference in frequencies, 
  and hence their synchrony is disrupted.
This effect also destroys 
the second synchrony ($c$-$g$ or $g$-$h$). 
Thus, 
the system returns 
to the original no-synchrony state.
In this way,  
the pairs of synchronization in the network D are never settled 
and the amplitudes of oscillators $c$, $f$, $g$ and $h$ 
continue to change spontaneously.
For larger networks, 
it becomes more difficult to determine the pairs of synchronizing 
oscillators in a consistent manner.
Thus, complete layered behaviors becomes rare in larger networks.


In summary, we have investigated networks of limit cycle oscillators 
the frequencies of which change widely along with the amplitude changes,
and report novel complex behaviors: 
the layering and the continuous large fluctuation of amplitudes.
In the analysis, 
we explained a mechanism that determines  the behaviors 
of amplitudes from the local structure of the network. 
The behavior of amplitude primarily depends on the number of 
connections the oscillator has,  
but also depends on the states of the connecting oscillators 
which are determined by their numbers of connections.
Therefore, the global structure of the network topology affects 
the behavior of each oscillator.

The dependence of the frequency 
on amplitude changes is key for a variety of phenomena.
In layered states, the differences in amplitudes originate from 
the differences in frequencies, and this  
assures the decoupling of oscillator interactions by averaging.
Thus, the differences in amplitudes are sustained.
The same effect enables the globally coupled oscillators 
to form stable multi-cluster states \cite{Dai06,TF07}. 
By contrast, in fluctuating states, amplitude changes 
result in switching
between decoupled and influential interactions.
This emergence and annihilation of effective interactions 
sustainably drives the fluctuation of amplitudes.

In this analysis, we also show that the topology 
of networks determines the behaviors of the oscillator networks,
and reveal how a small difference in topology 
brings about qualitative change in behaviors.
This indicates that 
exploring the detailed topology of oscillator networks \cite{GS02}
is as important  as analyzing them  statistically\cite{collect}.

This change in effective frequencies (time scales) and the associated 
change of effective interactions are important properties of 
complex systems.
Such behaviors are frequently observed in biological systems
as adaptations of time scales. 
For example, nerve systems are thought to code information 
by changing the firing frequencies of neurons as well as their phase 
relations \cite{BD04}. 
The complex dynamics of both properties are expected to 
work as unified systems \cite{Kay05}. 
These relations are also observed in 
adaptive behaviors of single cell organisms which contains  
a variety of time scales \cite{Shi95,Tak97}.
Our study extracts the mechanism of changing time scales
from a general limit cycle system, and 
reveals the origin of these complex phenomena.

\begin{acknowledgments}
The author is grateful to K. Kaneko, S. Ishihara, and K. Fujimoto 
for helpful suggestions.
\end{acknowledgments}

\end{document}